\documentclass[iop]{emulateapj}
\usepackage{verbatim}
\usepackage{rotating}
\usepackage{mathtools}
\usepackage{natbib}
\usepackage{todonotes}
\usepackage{float}

\begin{document}
\title{Inferring Gravitational Potentials from Mass Densities in Cluster-sized Halos}
\author{Christopher J. Miller\altaffilmark{1,2} , Alejo Stark\altaffilmark{1}, Daniel Gifford\altaffilmark{1}, Nicholas Kern\altaffilmark{1}}
\altaffiltext{1}{Department of Astronomy, University of Michigan, Ann Arbor, MI 48109 USA}
\altaffiltext{2}{Department of Physics, University of Michigan, Ann Arbor, MI 48109, USA}
\email{christoq@umich.edu}

\begin{abstract}
We use N-body simulations to quantify how the escape velocity in cluster-sized halos maps to the gravitational potential in a $\Lambda$CDM universe. Using spherical density-potential pairs and the Poisson equation, we find that the matter density inferred gravitational potential profile predicts the escape velocity profile to within a few percent accuracy for group and cluster-sized halos (10$^{13} < M_{200} < 10^{15}$M$_{\odot}$, with respect to the critical density.) The accuracy holds from just outside the core to beyond the virial radius. We show the importance of explicitly incorporating a cosmological constant when inferring the potential from the Poisson equation. We consider three density models and find that the Einasto and Gamma profiles provide a better joint estimate of the density and potential profiles than the Navarro, Frenk and White profile, which fails to accurately represent the escape velocity. For individual halos, the 1$\sigma$ scatter between the measured escape velocity and the density-inferred potential profile is small ($<$5\%). Finally, while the sub-halos show 15\% biases in their representation of the particle velocity dispersion profile, the sub-halo escape velocity profile matches the dark matter escape velocity profile to high accuracy with no evidence for velocity bias outside 0.4$r_{200}$. 
\end{abstract}


\section{Introduction}
\label{sec:introduction}
        Cosmological N-body simulations are a theoretical tool to understand how gravity in a dynamical space-time governs the formation of massive objects. Cluster-sized halos are recently-formed (if not {\it still forming}) objects, with sizes that reach beyond the scale of the effects from baryonic physics. In the cores of clusters where the baryonic cooling time is short, localized disturbances are not yet well understood, but researchers model them using hydrodynamics and astrophysical feedback mechanisms \cite[{\it e.g.,} ][]{Martizzi12, Martizzi14, Pike14}. From these simulations we have learned that  within cluster cores, baryonic physics can affect the local density and the gravitational potential and thus affect the dynamics of the tracers \citep{Lau10}. Outside cluster cores, it is only gravity and the expanding space-time which govern the potential and the dynamics.
        
        Clusters grow through infall and accretion \citep{vdBosch02,McBride09}. Particles and smaller sub-halos are held, captured, or released over time as the systems grow in mass and the gravitational potential deepens. Under Newtonian dynamics, the escape velocity is related to the gravitational potential of the system,
        \begin{equation}
            v_{esc}^2(r) = -2\phi(r) .
        \label{eq:escape}
        \end{equation}
        The extrema of the tracer velocities in the radius/velocity phase space define a surface which we call the escape velocity {\it edge}. In other words, particles, sub-halos, semi-analytic galaxies, etc, all exist in a well-defined region of the radius/velocity ($r-v$) phase-space bounded by a sharp escape velocity edge.  By determining this escape velocity surface, one is directly measuring the projected potential profile. 
        
        We can then use the Poisson equation to infer the mass density profile from the potential via
        \begin{equation}
        \nabla^2\phi(r) = 4\pi G \rho(r),
        \label{eq:poisson}
        \end{equation}
     where $G$ is the gravitational constant, $\phi$ is the gravitational potential, and $\rho$ is the matter density. In practice, the escape-edge is used to estimate $\phi$ and infer cluster masses \citep{Rines08,Rines13,Geller13,Gifford13b}. 
     
\citet{Gifford13a} (hereafter GM) find that when calibrated through an argument based on virial equilibrium, the escape velocity technique allows one to infer unbiased cluster-sized halo masses with low scatter ($\sim 10\%$) in three dimensional simulated data. The GM result suggests that the actual gravitational potential is precisely traced by the escape edge. Our primary goal for this paper is to test this hypothesis.
        
However, GM also showed that when using the \citet{Navarro97} (hereafter NFW) mass profile to predict the potential profile via the Poisson equation, the masses are biased low by $\sim 10\%$. \citet{Serra11} show that the NFW potential over-predicts the numerically evaluated potential by $>$ 10\%. This is consistent with the mass underestimation found by GM (see their equation 6) and relates to how the caustic technique is applied within the NFW formalism. \citet{Serra11} propose that the mass outside the cluster exerts a pull which would lower the numerical value of potential and explain the difference, but this is not a satisfactory explanation because the presence of mass would only increase the fractional difference between the numerical and the NFW-inferred potential profile. Another aim of this paper is to reconcile this reported discrepancy between the expected and actual accuracy of the escape-velocity technique as a representation of the gravitational potential in cluster-sized halos.

        When using the Poisson equation to infer the potential one needs to be concerned with the accuracy of the mass density profile. Because the potential is determined from an integration over the density to well beyond the virial radius, we require that the spherically averaged cluster density profile be reasonably accurate over a wide range of scales. However, even if the density profile is not entirely accurate, a steep drop-off in the density means that there is little mass in the outskirts to contribute to the deepening of the potential. 
        
For instance, while many authors have shown that the NFW is a good measure of the density profile outside the core to the virial radius \citep{Cole96,Tormen97,Bullock99}, it has a shallower outer profile than the Einasto profile \citep{Einasto69}. The Einasto profile is also a better model of the density profile \citep{Merritt06}. This motivates us to consider multiple density models when applying the Poisson equation to infer the potential.

In this paper, we focus on understanding the precision and accuracy of the gravitational potential as measured by the escape velocity surface for individual cluster-sized halos in N-body simulations. We utilize observables that can in principal be measured, such as the density profile (e.g., through gravitational lensing) and the phase-space edge (e.g., through spectroscopic surveys). We make extensive use of parametric models of the density and potential via the Poisson equation (Equation \ref{eq:poisson}). We focus only on 3D information in this paper, leaving the challenges of projected measurements and experiment-specific configurations to other efforts \citep{Gifford16,Gifford13b}.
\begin{figure*}
\plotone{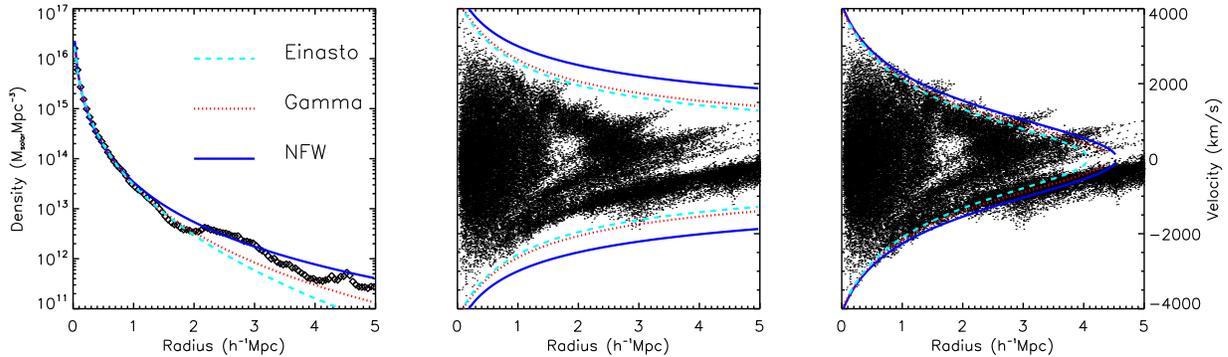}
\caption{{\bf Left:} The spherically averaged density profile of a halo from the Millennium Simulation (M$_{200} = 6.3\times10^{14}$M$_{\odot}$ and r$_{200} = 1.34$ Mpc at 200$\times$ the critical density). The three lines are fits to the density profile (squares) over the range $0 \le r_{200} \le 1$ using Equations \ref{eq:nfw_den},\ref{eq:gamma_den}, and \ref{eq:einasto_den}. {\bf Middle and Right:} The radius-velocity phase-space of the particles. These are the radial components of the particle velocities and include the Hubble flow. The lines in the {\bf middle and right} panels are the predicted escape velocity profile from the Poisson equation and the fits to the density profiles (Equations  \ref{eq:nfw_pot},\ref{eq:gamma_pot}, and \ref{eq:einasto_pot}). In the {\bf middle} panel, the Newtonian potential is integrated to infinity. In the {\bf right} panel the $\Lambda$CDM potential is integrated to the radius r$_{eq}$ where the gravitational force from the halo balances the expansion of the space in a $\Lambda$CDM universe. No dynamical information from the particles is used in the prediction of the escape-edge in the middle and right panels. }
\label{fig:dens_pot_pair}
\end{figure*}   
    
\section{Theory}
\label{sec:theory}

             Consider a mass distribution described by a spherical profile such that the mass density $\rho$ and the potential $\phi$ radial profiles are related by the Poisson equation (\ref{eq:poisson}):
        \begin{equation}
        \phi(r) =-{\rm 4\pi G}\Big{[} \frac{1}{r}\int_{0}^{r}\rho(r')r'^2dr' + \int_{r}^{\infty}\rho(r')r'dr'\Big{]}.
        \label{eq:phi_poisson}
        \end{equation}
        
         Equation \ref{eq:phi_poisson} allows one to analytically calculate the potential profile $\phi$ for spherical density models in a static universe and for isolated systems.

\subsection{Analytical Density-Potential Pairs}
        
        There exist analytic formulae which have been shown to fit the density profiles of halos in N-body simulations.  We consider the following three: the NFW profile, the Gamma profile \citep{Dehnen93}, and the Einasto profile \citep{Einasto69, Retana12}. Using equation \ref{eq:phi_poisson}, we have:

        \begin{subequations}\label{eq:nfw}
         \begin{align}
          \rho(r) &= \frac{\rho_0}{(r/r_0)(1+r/r_0)^2}  \label{eq:nfw_den} \\ 
          \phi(r) &= -\frac{{\rm 4\pi G} \rho_0 (r_0)^2 \ln(1+r/r_0)}{r/r_0} \label{eq:nfw_pot}
         \end{align}
        \end{subequations}
    
        \begin{subequations}\label{eq:gamma}
         \begin{align}
          \rho(r) &=  \frac{(3-n) {\rm M}}{{\rm 4\pi}}\frac{r_0}{r^n}\frac{1}{(r+r_0)^{4-n}} \label{eq:gamma_den} \\
          \phi(r) &= \frac{{\rm GM}}{r_0}\frac{-1}{2-n}\Big{[}1-\Big{(}\frac{r}{r+r_0}\Big{)}^{2-n}\Big{]},  n \ne 2 \label{eq:gamma_pot} \\
          &= \frac{{\rm GM}}{r_0}{\rm ln}\frac{r}{r+r_0},  n=2 \nonumber
         \end{align}
        \end{subequations}
        
        \begin{subequations}\label{eq:einasto}
         \begin{align}
          \rho(r) &=  \rho_0 \rm{exp} \Big{[}-\Big{(}\frac{r}{r_0}\Big{)}^{1/n}\Big{]} \label{eq:einasto_den}\\
          \phi(r) &= \frac{{\rm -GM}}{r} \Big{[} 1 - \frac{\Gamma\big{(}3n,\frac{r}{r_0}^{(1/n)}\big{)}}{\Gamma(3n)} + \frac{r}{r_0}\frac{\Gamma\big{(}2n,\frac{r}{r_0}^{(1/n)}\big{)}}{\Gamma(3n)}\Big{]} \label{eq:einasto_pot}
         \end{align}
        \end{subequations}
        where $\rho_0$ or $\rm{M}$ is the normalization, $r_0$ is the scale radius, and $n$ is the index. Equations \ref{eq:nfw}, \ref{eq:gamma} and \ref{eq:einasto} are examples of density - potential \emph{pairs} which share the same values for the shape parameters in the radial profiles of both the density and the potential. In other words, given a fit to the spherical density profile, one can infer the shape of the gravitational potential through these equations.

        In Figure \ref{fig:dens_pot_pair} we show an example halo with M$_{200} = 6.3\times10^{14}$M$_{\odot}$ and r$_{200} = 1.34$ Mpc from the Millennium Simulation \citep{Springel05}. We use radii and masses with respect to 200$\times$ the critical density throughout. The left panel shows the spherically averaged density profile and the three model fits from Equations \ref{eq:nfw_den}, \ref{eq:gamma_den}, and \ref{eq:einasto_den}. The models are fit over the range $0.0 \le r/r_{200} \le 1$. While the models are nearly identical within $r_{200}$, they differ significantly in the outskirts. We note that this is a single halo and is meant to illustrate the model differences. A statistical analysis is conducted in Section \ref{sec:comparison}.
        
        The middle panel shows the radius/velocity phase space of the particles within this halo. We use the radial components of the velocities of each particle and include the Hubble flow in the velocities. Notice that the particle edge contains a fair amount of localized structure due to infall. This cluster is dynamically active. The lines in the middle panel of Figure \ref{fig:dens_pot_pair} show the predicted escape velocity edge for the three models using the Poisson equation and the fits to the density profiles and using Equations \ref{eq:nfw_pot}, \ref{eq:gamma_pot} and \ref{eq:einasto_pot}. We consider each of the three models separately and infer model parameters by minimizing the $\chi^2$ difference to the density profiles. Notice that the naive use of the Poisson equation and equation \ref{eq:escape} over-predicts the escape edge.
     
\subsection{Integration Limit on $\phi$ in a $\Lambda$CDM universe}

      In the middle panel of Figure \ref{fig:dens_pot_pair}, the density-inferred escape velocities use the simple Newtonian formalism (equation \ref{eq:escape}) and an integration radius that requires escape to infinity. Equation \ref{eq:phi_poisson} ignores the added potential term from the cosmological constant, $\Lambda$. As shown in \citet{Behroozi13}, there exists a radius r$_{eq}$ where the radial inward pull from gravity balances the radial outward pull of the expanding universe. This radius can be derived in a simple way using only the radial components of the tracer velocities: $r_{eq}^3 = -GM/qH^2$, where $H$ is the Hubble parameter and $q$ is the deceleration parameter, $\Omega_m$/2 - $\Omega_{\Lambda}$. Behroozi et al. also consider non-radial motion, but in this work, we focus only on the radial component for both the theory and the measured velocities.
 
      We now revisit the limits of the integral on Equation \ref{eq:phi_poisson} and require tracers to escape only to $r_{eq}$. We also include the $\Lambda$-term and apply the same integration limit to the effective potential (Equation \ref{eq:phi_lambda}). Following Behroozi et al. (2013), the radial component of the escape velocity in an accelerating universe should be:
       \begin{equation}
      \Phi = \frac{v^2_{esc}}{2} = -(\phi(r) - \phi(r_{eq})) - \frac{qH^2}{2} (r^2 - r_{eq}^2). \label{eq:phi_lambda}
      \end{equation}
A similar derivation is provided by \cite{Behroozi13} for a point source\footnote{Note the sign difference compared to Behroozi et al. (2013), where we use the classical definition such that q is a deceleration.} In other words, the radial component of the escape velocity is zero (relative to the cluster) at $r_{eq}$, where a tracer is then picked up by the expanding universe. Equation \ref{eq:phi_lambda} means that the escape speed from a galaxy cluster in a $\Lambda$CDM universe is less than the Newtonian escape speed (equation  \ref{eq:escape}.)

We show the revised $\Lambda$CDM-specific escape velocity profile using Equation \ref{eq:phi_lambda} in the right panel of Figure \ref{fig:dens_pot_pair}. Notice that both the shape and the amplitude of the predicted escape edge match the phase-space edge using the tracer particles. We note that Equation \ref{eq:phi_lambda} utilizes the definition of $r_{eq}$ derived for a point mass, such that we require the entire cluster mass to be contained within $r_{eq}$. 
\begin{figure}
\plotone{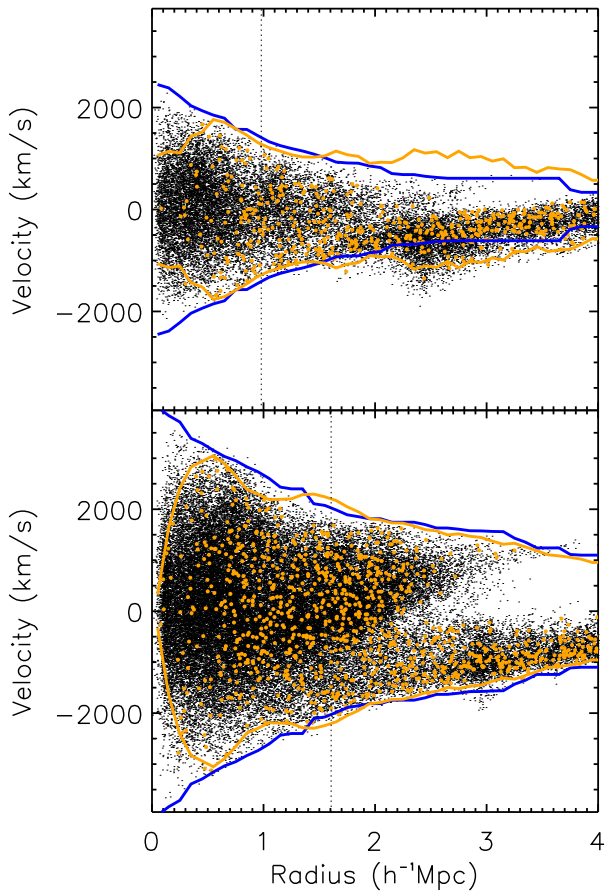}
\caption{The radius-velocity phase spaces of a low mass (top- $2.4\times10^{14}$M$_{\odot}$) and high mass (bottom- $1.0\times10^{15}$M$_{\odot}$) cluster in the Millennium simulation. The dots are particle radial positions and radial velocities. The orange circles are sub-halo radial positions and radial velocities. The lines are the measured escape edges for the particles (blue) or the sub-halos (orange). Notice the increasing statistical bias in the sub-halo edges compared to the dark matter edges towards the core where the sampling is low. The dotted vertical bar is the location of $r_{200}$.}
\label{fig:edges}
\end{figure}  

\section{Analysis}
\label{sec:analysis}

In the previous section, we showed qualitatively that we can predict the potential profile, and thus the phase-space escape velocity edge profile, from the density profile alone. We showed a few clusters to highlight the the theoretical expectations. However, in order to quantify how well the measured escape velocity profiles match the profiles predicted from the density-inferred potential, we need to measure the phase-space escape edges. We also need to conduct the analysis over a larger sample of clusters in the simulations. 
   
\subsection{Measuring the Phase-space Edges}
\label{sec:edges}    

We follow \citet{Diaferio97}, where the edges are defined by the minimum of the two maxima in the positive and negative velocity sectors of the radially binned phase-spaces. Beyond the core (0.2h$^{-1}$Mpc) we enforce each min/max edge to be equal to or lower than the previous edge, such that the edge profiles are monotonically decreasing. This is slightly different than \citet{Diaferio97}, who use an additional free parameter to limit the radial up and down variations in the edge profile in order to mitigate the effects from local structure in the phase-spaces (see Section \ref{sec:introduction}).

In Figure \ref{fig:edges} we show the measured edge for two halos of different mass. The velocities in the phase-spaces are the radial components of the tracers and include the Hubble expansion with respect to the cluster. We also measure the edges using the sub-halos identified within the main halos (orange circles and lines). We discuss the sub-halos in Section \ref{sec:subhalos}.

\subsection{Predicting the Escape Edges}
\label{sec:comparison}

      Our next goal is to quantify how well the predictions of the $\Lambda$CDM potential, using Equations \ref{eq:nfw}, \ref{eq:gamma}, and \ref{eq:einasto} and exemplified in Figure \ref{fig:dens_pot_pair}, compare to the measured escape velocity edge, exemplified in Figure \ref{fig:edges}.

     We use the 100 halos from the Millennium Simulation and their particle data as described in \citet{Gifford13b}, which have a uniform mass sampling from $1\times10^{14} \le M_{\odot} \le 2\times10^{15}$. We work in physical units of km/s, i.e., $\sqrt{-2\phi}$. For the remaining analyses we follow the same procedures. First, we fit the spherical radial density profiles to each of the three models and for each halo separately. We then take the best-fit density model parameters to make a prediction of the escape velocity edge. We then compare the predicted escape velocity profile to the measured phase-space edge. We measure the accuracy using a radial average of the fractional differences between the predicted and the measured escape surfaces. We do the same for the scatter, which is determined using all 100 halos.

\begin{figure*}
\plottwo{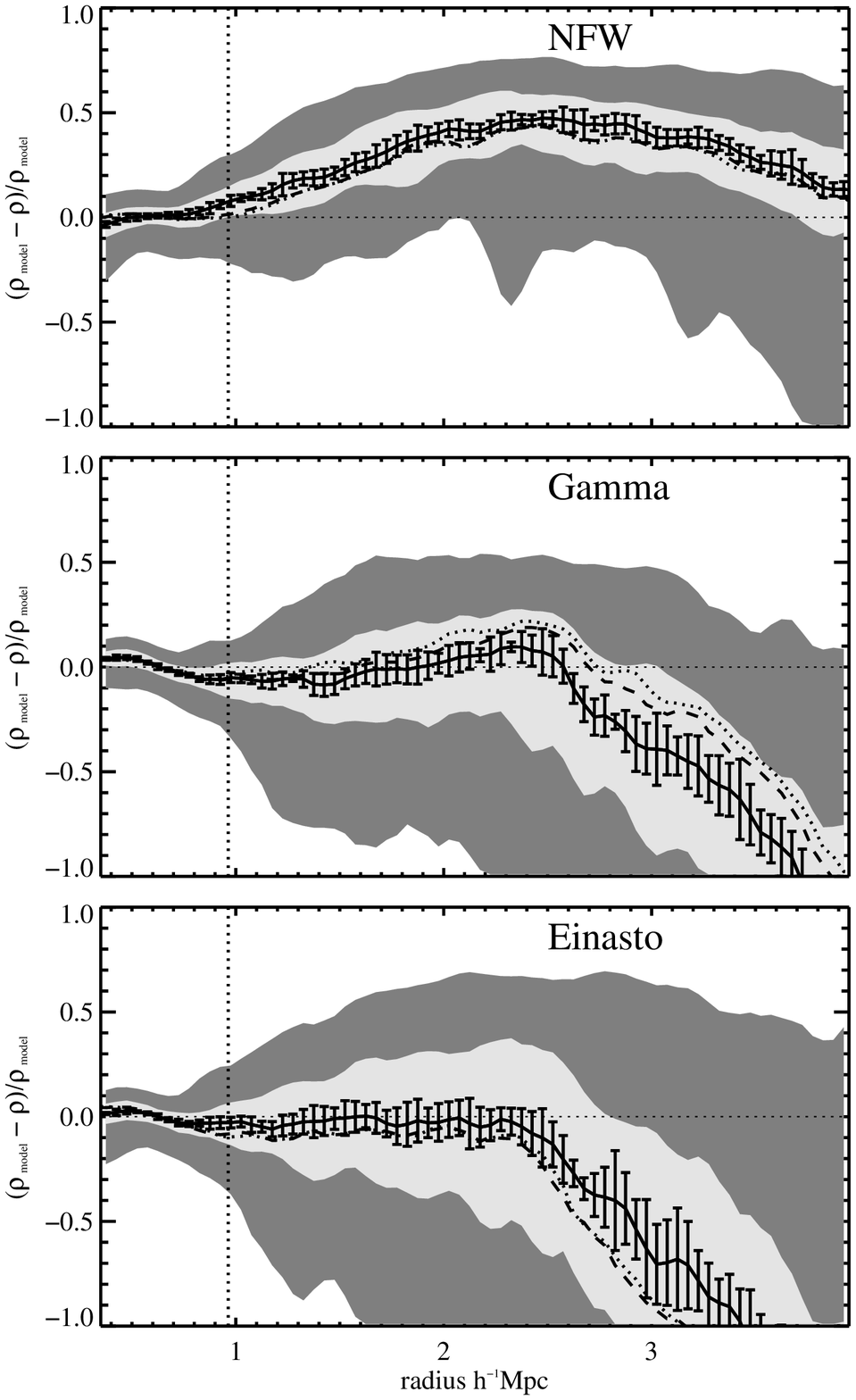}{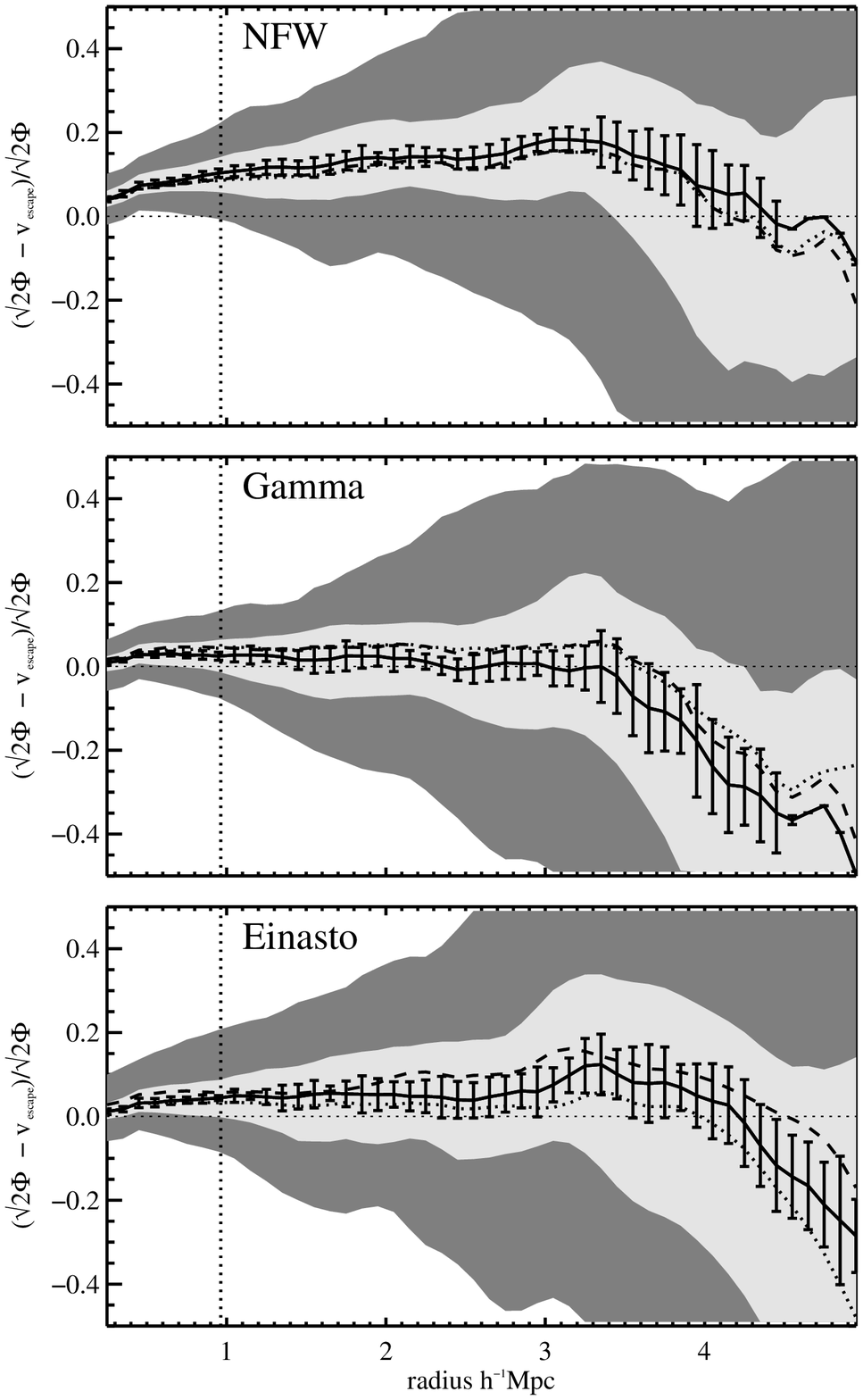}
\caption{The fractional difference between theory and simulation observables for the NFW potential-density Poisson pair in cluster-sided halos. {\bf Left} shows the fractional difference between the particle density profiles and the model profile fits. The median of the 100 halos is the solid line and the error bars are determined from boot-strap re-sampling of the median. The dark grey band encompasses 90\% of the individual halo profiles and the light grey band 67\%. {\bf Right} shows the fractional difference between the measured $v_{esc}$ edges and the inferred $\Lambda$CDM potential ($\sqrt{-2\Phi}$) based on the best model fits to the density profiles.  The individual profiles are determined relative to the average particle velocities and positions (solid) and only small differences appear when we use the sub-halo positions or re-define the particle mean velocity and position using only particles within 0.5 r$_{200}$ (dashed, dotted). Note that the NFW model density profile over-predicts the measured density from $1 \le r_{200} \le 4\times$ r$_{200}$. As a consequence, an over-abundance of mass is integrated into the NFW potential profile, thus inflating the expected potential compared to the measured escape velocity profile. The Gamma and Einasto fits to the density profiles provide a more accurate representation of their respective potential profiles.} 
\label{fig:denpotpair}
\end{figure*}

      In Figure \ref{fig:denpotpair}, we show fractional differences between the model and the data. We  calculate errors on the median (solid line) using bootstrap re-sampling with replacement. We also show the cluster-to-cluster scatter as the light and dark gray bands (67\% and 90\% respectively).

      We find that all of the profiles perform well when measuring the density utilizing all particles within the range $0.3 \le r/r_{200} \le 1$. Beyond $r_{200}$ it is clear that the Gamma and Einasto density profiles fall off much more quickly compared to the NFW. We also find that the Einasto and Gamma potential profiles perform better than the NFW when using the density profile to predict the escape edge. The NFW predicts an escape edge that is biased high at a level of 10-15\% out to a few times $r_{200}$. This is due to the fact that the density profile is over-estimated out to 4$\times$ $r_{200}$. On the other hand, the Gamma and Einasto density profiles are more accurate than the NFW beyond $r_{200}$ and drop off quickly to produce potential profiles that are nearly unbiased ($\sim$ 3\% or less) out to 3$\times$ $r_{200}$. Regardless of the model, the cluster-cluster variation (or scatter) between the predicted and measured escape edges is  $< 5\%$ percent ($< r_{200}$) for most clusters.

      Figure \ref{fig:denpotpair} also shows what happens when we move the center of the halo. The solid line uses halo centers defined by the mean position and velocity of all particles within $r_{200}$. The dotted line uses the position of the central halo defined by SUBFIND \citep{Springel01}. The dashed line uses the mean position of all particles within 0.5$r_{200}$.  The Einasto profile is most sensitive to the positional choice of the main halo, whereas the Gamma and NFW profiles show the least variation (the differences are hardly noticeable in Figure \ref{fig:denpotpair}).

      We conclude that the joint accuracy and precision of the density and the potential depends on the choice of the parametrized density model used. Both the Gamma and the Einasto profiles produce nearly unbiased density-potential pairs when compared to observables, while the Einasto profiles are most sensitive to how the halo centers are defined. As noted in the Introduction, \citet{Serra11} attribute the lower numerically integrated potential (equation \ref{eq:phi_poisson}) compared to the NFW potential (equation \ref{eq:nfw_pot}) to mass outside the cluster that is not accounted for by the NFW profile shape. We show that the opposite is true. The NFW model density profile over predicts the true density profile outside $r_{200}$ and thus over predicts the true gravitational potential via the Poisson equation by 10-15\%.

\subsection{Mass and Redshift Dependence}
Next, we examine how the edge varies as a function of halo mass and redshift. In this case, we use a new sub-set of the Millennium simulation with the 100 most massive halos halos smaller than $1\times10^{14}$M$_{\odot}$. The minimum mass of this new subset is $\sim 1\times10^{13}$M$_{\odot}$, i.e.  a factor of 10 smaller than the previous sample. 

We measure the spherically averaged density profiles and infer the escape edges via the potential from Equation \ref{eq:einasto}. The edges are measured using the same algorithm as applied to the more massive subset studied previously. We find no statistical or systematic difference between the high mass and low mass halo datasets. The density profile predicts the escape edge via the Poisson equation to the same level of accuracy and precision for over two orders of magnitude in cluster halo mass.

\begin{figure}
\plotone{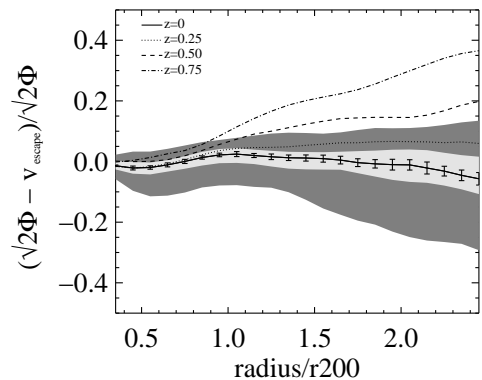}
\caption{The variation in the  fractional difference between the measured $v_{esc}$ edge and the predictions from the density profile from redshift 0 to 0.75. The light (dark) grey bands represent the 67\% (90\%) scatter of the individual halos.}
\label{fig:versusz}
\end{figure}

We study the low mass clusters at four different simulation snapshot outputs, corresponding to $z=0, 0.25, 0.5$, and $0.75$. We keep the 100 most massive halos in each snapshot as provided by the ``millimil'' subset of the Millennium data.  At $z > 0.75$, the deceleration parameter goes from negative to positive. As the cluster density profiles evolve with redshift, we fit the profiles separately for each halo at each redshift. Instead of physical coordinates, we use radial coordinates with respect to the $r_{200}$ of each cluster for the profiles, due to the fact that the cluster sizes also evolve with redshift. In Figure \ref{fig:versusz} we show that within $r_{200}$, the edge can be accurately predicted from the density and Poisson equation to z=0.75. However outside the virial radius, the edge is increasingly under-predicted compared to the model, as a function of increasing redshift. We find the same result when using the mean background density as opposed to the critical background density when defining the cluster masses and radii. 

We can explain this by the growing influence of the halo gravitational potential well on the dynamics of the infall regions around the clusters. Over time, the measured escape edge in the outskirts of galaxy clusters grows in amplitude to represent the predicted escape velocity defined by the potential. This dynamical evolution of the escape edge in the cluster infall regions is important for studies which use the escape velocity technique to measure mass profiles to well beyond the virial radius \citep{Rines06,Rines13}.

\subsection{Particles vs. Subhalos}
\label{sec:subhalos}

Having defined the baseline accuracy and precision of the escape velocity technique for cluster-sized halos using the particles, we ask whether other tracers of cluster potential can be used. We use resolved sub-halos defined for the Millennium simulation by SUBFIND \citep{Springel01}.  As an example, the sub-halos for two clusters are shown as the orange circles in Figure \ref{fig:edges}. There are two important issues with the sub-halos that are evident in this figure. First, the sub-halos decrease in density towards the core while the particles increase in density. Second, the sub-halos do not track the phase-space near $\Delta v = 0$ within $r_{200}$. Both of these effects are the result of sub-halo destruction through gravitational interactions with the density field: galaxies would not be destroyed so easily. However, by using only the sub-halos which have survived mergers as a tracer of the particle phase-space, one is weighting the velocity distribution in an unfair way with respect to both the dark matter particles as well as any realistic galaxy populations. 

\cite{Wu13} review the current consensus on velocity bias in simulated halos. As measured by the velocity dispersion, Wu et al. find that sub-halos typically show 10-15\% positive biases \cite[{see also}][]{Lau10}. This is a manifestation of how the radius/velocity phase-space is sampled by the resolved sub-halos. One can draw from the phase-space in any number of ways, any of which may show positive or negative biases compared to the full representation of the phase-space by the particles. In the case of sub-halos, they can easily be destroyed through interactions causing a paucity of tracers with low velocities. The end result is a sub-halo velocity dispersion that is biased with respect to the particles.

\cite{Gifford13b} showed that the virial masses and the caustic masses of halos in the Millennium simulation are biased high when using only the sub-halos (by $\sim$ 35\% and 30\% respectively for well-sampled phase-spaces). These biases are always a result of the velocity dispersion. The virial mass is biased simply because it is directly related to the velocity dispersion \citep{Evrard08}. The caustic mass is biased because the standard ``caustic'' technique calibrates the escape edge to the velocity dispersion \citep{Diaferio99,Gifford13a}. In this work, we do not calibrate the escape surface according to virial equilibrium, but we measure it directly. Therefore, we can test  whether the sub-halos are in fact biased tracers of the escape-edge by comparing to the particle edges. 

First, we need to separate systematic velocity biases (i.e., along the vertical axis of the phase-space diagrams) from statistical sampling biases (i.e., along the horizontal axis). While most of our halos have thousands of particles in each radial bin of the phase-space, there are only tens of sub-halos in any bin. This can cause a sampling bias as a function of radius due to the small number of objects per bin. This bias is purely statistical and is visible in Figure \ref{fig:edges}. We can determine the level of this bias by sub-sampling from the particles to match the number of sub-halos. We use 100 uniformly random sub-samples of the particles per bin per cluster. We then calculate the difference between the sub-sampled edge and the full particle edge. Not surprisingly, we find a statistical sampling bias that gets worse as we move into the core of the clusters and the sub-halo density relative to the particle density decreases. We calculate the radial difference between the full and sub-sampled edges with respect to the particle edge as determined beyond $r_{200} = 3h^{-1}$Mpc, well beyond the radius where sub-halo interactions are common. We then apply this statistical sampling correction to the measured sub-halo escape edges. We note that the sampling bias results in an escape edge that is biased low and in the opposite direction of the halo bias reported in \citet{Lau10}, \citet{Gifford13b} and \citet{Wu13}.

\begin{figure}
\plotone{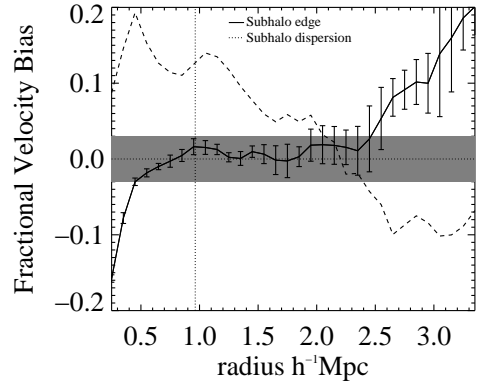}
\caption{The fractional sub-halo velocity bias profile with respect to the particles for the velocity dispersion and the edge. In both cases, we remove the statistical bias with results from the low sampling of the sub-halo population. While the sub-halos have a biased velocity dispersion with respect to the dark matter particles, the escape edge is well-constrained by the sub-halos. The vertical line is the average $r_{200}$ for the sample. The gray band represents the $\pm{3}\%$ scatter on how well the density-inferred potential predicts the measured escape edge from the particles.
}
\label{fig:velbias}
\end{figure}

In Figure \ref{fig:velbias}, we show the sampling corrected sub-halo velocity dispersion and edge bias determined as the fractional difference from the particles. To ensure a fair comparison to the velocity dispersion, we apply the same sampling correction procedure as we did for the escape edges. Notice that the velocity dispersion profile shows positive biases $\sim$ 15\%, identical to what is presented in \cite{Gifford13b}. However, the escape-edge based on the sub-halos is unbiased beyond $\sim$ 0.4 h$^{-1}$Mpc to at least $\sim$ 2$\times r_{200}$.

The fact that the sub-halo edge is unbiased is an important  result. The edge is a well-defined and sharp feature of the phase-spaces of halos in simulations and so long as the sampling is high enough, the edge will be detected regardless of how the sampling is done. Gravity insists that there can be no population of tracers which exist above the escape edge.  We note that there can be sub-halos which momentarily live above the edge while they are escaping (see the top panel of Figure \ref{fig:edges}), but these are rare and fleeting and do not systematically bias the edges for all halos over all radii.

\subsection{Non-Radial Escape}
\label{sec:nonradial}

Throughout this work we focus on radial escape. However, it is known that escape along tangential orbits requires more kinetic energy than radial escape  \citep{Behroozi13}. Therefore, we investigate the escape edge as measured through the velocity vectors along the $\theta$ and $\phi$ directions. We treat the analysis of the non-radial motion identically to the radial motion and identify the edge as described in Section \ref{sec:edges}. The only difference is that the phase-spaces utilize particle velocities tangential to the sphere. We take $v_{esc}$(non-radial) as ($v_{esc}(\theta)$  +$v_{esc}(\phi)$)/2.

\begin{figure}
\plotone{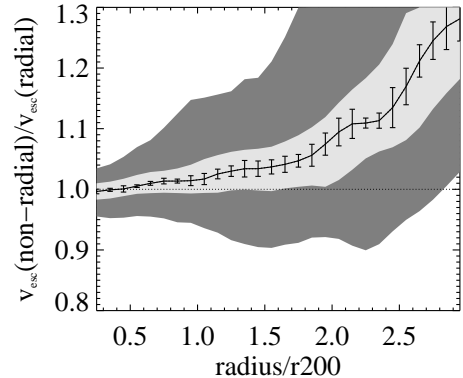}
\caption{The fractional difference between the radial and non-radial components of the velocities that comprise the escape edge. Tangential motion increases the escape edge compared to radial motion. The light (dark) grey bands represent the 67\% (90\%) scatter of the individual halos.}
\label{fig:nonradial}
\end{figure}

In Figure \ref{fig:nonradial} we show the fraction of the radial versus the non-radial components of the velocity in the escape edge. As expected, we see that tangential components grow with respect to the radial component with increasing radius. Within the virial radius, the fractional difference is small (a few percent) growing to a $>$10\% at a few virial radii. 

Figure \ref{fig:nonradial} is also a representation of the velocity anisotropy of the particles which comprise the edge. Notice that the edge is nearly isotropic, such that radial and non-radial components of the edge velocities are nearly equal. 

\section{Summary}

We quantified the density-inferred gravitational potential compared to the escape velocity surface for individual cluster-sized halos in N-body simulations. Throughout, we utilized observables, such as the density profile (e.g., through gravitational lensing) and the phase-space edge (e.g., through spectroscopic surveys). We then applied the Poisson equation on potential-density pairs to predict the potential and thus the escape velocity profile of cluster-sized halos. Our main conclusions are:
\begin{itemize}
\item{The upper limit on the integral over the density in the Poisson equation needs to be physically meaningful and must incorporate the added potential from the cosmological constant. Specifically, we find that particles and sub-halos are escaping to the radius at which the gravitational force on a tracer is balanced by the pull of the expanding universe.}
\item{The Einasto and Gamma density profiles can predict the escape edge of the radius-velocity phase-space to within 3\% accuracy and 5\% precision from outside the core to $\sim 3$ virial radii for low and high mass clusters. Within the virial radius, this precision and accuracy holds to $z = 0.75$.}
\item{The NFW profile over-predicts the halo density profile beyond $r_{200}$ and thus the potential profile at all radii by 10-15\%. In other words, the NFW model is not a true potential-density pair in the context of the Poisson equation.}
\item{The sub-halo velocity dispersion profile is biased high compared to the dark matter particles by as much as 15\%. However, the sub-halo escape velocities trace the dark matter escape edge to high accuracy outside the core.}
\end{itemize}

We conclude that the Poisson equation for clusters in a $\Lambda$CDM universe results in a well-defined phase-space edge for the particles and sub-halos. The density profile alone can be used to predict the dynamically-inferred potential of groups and clusters. These results are informative and encouraging for mass estimation techniques based on the dynamical potential of clusters. 

In this work we utilize the 3-dimensional positions and velocities of the tracers to match the radial escape velocity to its prediction. However, the real universe is subject to  projection effects and line-of-sight observables, both of which smear the edge. Fortunately, it has been shown that stacked phase-spaces can recover this sharp phase-space caustic even for poorly sampled individual phase-spaces \citep{Gifford16}. By using stacked phase-spaces and stacked weak-lensing mass profiles, one can use current data to explore the very precise agreement expected for density-inferred escape edges. The methods discussed here have recently been applied to make predictions on how well phase-space edges can constrain modifications to gravity in the local Universe \citep{Stark16}.

We also report an important implication regarding the wide range of halo density models discussed in the literature. Previous research has focused on the inner core regions of clusters when identifying differences between universal density profiles \cite[{\it e.g.,} ][]{Merritt06,Diemer15}. Our work does not focus on which of the NFW, Gamma, or Einasto-shaped profiles are the best when measuring the density. Instead, we focus on the joint recovery of the density and potential profiles. The density profile beyond $r_{200}$ plays an important role in the accuracy of the predicted the phase-space edge. It is possible that there is an even more accurate functional form which describes the Poisson-pair profiles of cluster-sized halos.

\section{Acknowledgements}
The authors want to thank Jessica Kellar and August Evrard for their helpful comments and discussion. We also thank the anonymous referee who made useful suggestions to improve the manuscript. This material is based upon work supported by the National Science Foundation under Grant No. 1311820. The Millennium Simulation databases used in this paper and the web application providing online access to them were constructed as part of the activities of the German Astrophysical Virtual Observatory (GAVO). The authors want to especially thank Gerard Lemson for his assistance and access to the particle data. 


\end{document}